\documentclass[pra,twocolumn,showpacs,preprintnumbers,amsmath,amssymb]{revtex4}
\newtheorem{theorem}{Theorem}[section]

\newtheorem{proposition}[theorem]{Proposition}

\newtheorem{lemma}[theorem]{Lemma}

\begin{document}

\title{Checking $2 \times M$ separability via semidefinite programming}
\author{Hugo J. Woerdeman}
\affiliation{Department of Mathematics,
The College of William and Mary,
Williamsburg, VA 23187-8795}
\email{hugo@math.wm.edu}

\begin{abstract}
In this paper we propose a sequence of tests which
gives a definitive test for checking
$2\times M$ separability.
The test is definitive
in the sense that each test corresponds to checking membership in
a cone, and that the closure of the union of all these cones
consists exactly of {\it all} $2 \times M$
separable states. Membership in each single cone
may be checked via semidefinite programming, and is thus a
tractable problem.
This sequential test comes
about by considering the dual problem, the characterization of all
positive maps acting ${\mathbb C}^{2 \times 2} \to {\mathbb C}^{M\times M}$. The latter
in turn is solved by characterizing
all positive quadratic matrix polynomials in a complex variable.
\end{abstract}

\maketitle

\section{\label{sec:level1}Introduction}

In the last decade entanglement
has been recognized to be a fundamental notion in
quantum information processing.
While initially entanglement lead to ``paradoxes''
in quantum mechanics, it was later discovered to be a
useful tool in teleportation, secure key distribution (quantum cryptography),
quantum computation, etc. (see, e.g., \cite{NC} and
\cite{BC}). Despite its importance there does not yet exist a tractable
conclusive test to check for entanglement or the lack thereof (separability).
There are several partial tests the most famous of which
is the Peres test \cite{P} which says that a separable state
remains positive under taking partial transposes. In some cases the
Peres test is also conclusive, namely in the low
dimensional cases $2\times M$,
$M=2,3$ (see \cite{HHH}) and in cases where the density matrix is of low rank
(see \cite{Horo}).
Recently, in \cite{DPS} the partial positive test
was combined with an extendibility property
to obtain a sequence of tests stronger than the Peres test, that all
separable matrices need to pass.
It is not known whether failure to pass this sequence of tests
yields entanglement.
Our main result concerns a sequence of tests that for the
$2\times M$ separability problem do the opposite.
Namely, passing one of the tests yields $2\times M$ separability.
This sequence of tests yields a definite answer to the
$2\times M$ separability problem
in the sense that any $2\times M$ separable matrix is arbitrarily
close to a matrix satisfying one of the tests.
Let us present the details.

Let $M_p(K)$ denote $p\times p$ matrices whose entries belong to $K$.
Recall that
a matrix $\rho \in M_2(M_M({\mathbb C}))$ is called $2\times M$ {\it separable}
if there exists $x_i \in {\mathbb C}^2$ and $y_i \in {\mathbb C}^M$ so that
$$ \begin{pmatrix} A & B \cr B^\dag & D \end{pmatrix} = \sum_{i} x_ix_i^\dag \otimes
y_i y_i^\dag . $$
Here $\ ^\dag$ denotes the conjugate transpose and
$\otimes$ denotes the Kronecker product.
Let $n \ge 1$ and
$\rho = \begin{pmatrix} A & B \cr B^\dag & D \end{pmatrix}$, where
$A, B,$ and $D$ are $M\times M$ matrices.
We introduce the following convex set, which may be empty.
Let
${\mathcal G} (\rho ; n ) \subset M_{n+1}(M_2(M_M({\mathbb C})))$
consist of all positive semidefinite block matrices
$\Gamma =(\Gamma_{ij})_{i,j=0}^n $ (notation: $\Gamma \ge 0 $) where
$$ \Gamma_{ij} = \begin{pmatrix}  \Gamma_{ij}^{(a)} & \Gamma_{ij}^{(b)} \cr
\Gamma_{ij}^{(c)} &  \Gamma_{ij}^{(d)} \end{pmatrix} \in
M_2(M_M({\mathbb C})) , 0 \le i,j \le n , $$
satisfy the conditions
\begin{itemize}
\item[(i)] $\sum_{i=0}^n \Gamma_{ii}^{(a)} = A$, \item[(ii)]
$\sum_{i=0}^n \Gamma_{ii}^{(d)} = D$, \item[(iii)]
$\sum_{i=0}^{n-1} \frac{1}{2} (\Gamma_{i,i+1}^{(c)} +
\Gamma_{i,i+1}^{(b)}) = B$, \item[(iv)] $\sum_{i=0}^{n}
(\Gamma_{i,i}^{(c)} - \Gamma_{i,i}^{(b)}) = 0$, and \item[(v)]
$\sum_{i=0}^{n-k} (\Gamma_{i,i+k}^{(c)} - \Gamma_{i,i+k}^{(b)})=
0$, $k=1,\ldots , n $.
\end{itemize}
We will be interested in matrices $\rho$ for which ${\mathcal G} (\rho ; n )$
is nonempty.
Therefore we define
$$ {\mathcal A}_n := \{ \rho \in M_2(M_M({\mathbb C})) : {\mathcal G} (\rho ; n ) \neq \emptyset \} . $$
It is straightforward to check that ${\mathcal A}_n$ is a closed convex cone.
Moreover, since ${\mathcal G} (\rho ; n ) \oplus 0_{2M}
\subseteq {\mathcal G} (\rho ; n+1 )$ one obtains
that ${\mathcal A}_n \subset {\mathcal A}_{n+1}$, $n \ge 1$.

It is an important feature of the set ${\mathcal G} (\rho ; n )$ that it is the intersection
of the cone of positive semidefinite matrices (PSD) and an affine set. As such,
the question of whether the set is empty or not, falls into a class of
well studied problems, called semidefinite programming (SDP). Checking
the nonemptyness of such an intersection is called the feasibility problem in
SDP, and several software packages (e.g., \cite{Overtonetal}, \cite{Boydetal})
are readily available to solve the feasibility problem numerically.
A good overview article on SDP is \cite{VB}.

We now state the main result.

\begin{theorem} \label{main}
The matrix $\rho \in M_2(M_M({\mathbb C}))$ is
$2\times M$ separable if and only if
$$
\rho \in \overline{\cup_{n\ge 1}
{\mathcal A}_n } . $$
\end{theorem}

Based on the above theorem one may now formulate the following algorithm for
checking for $2\times M$ separability. Given a time limit in which the check
needs to be performed, make $n$ as large as possible so that
$\rho \in {\mathcal A}_n$ may be checked via SDP within the given time limit.
In case the test comes out affirmatively, the given matrix $\rho$ is
$2\times M$ separable. When the test shows that $\rho \not\in {\mathcal A}_n$
the test is inconclusive. Still, given the content of Theorem \ref{main},
the negative outcome may be an indication that the matrix is entangled.
Of course, the larger $n$ was chosen, the stronger the indication is.

This is certainly not the first instance where SDP has been used for a
quantum information problem. In fact, the earlier mentioned
tests in \cite {DPS} may be done by SDP. Moreover, in
the papers \cite{AdM}, \cite{Rains}, \cite{Jezek}, \cite{Eldar}
connections have been made between several
other problems in quantum information and the versatile
SDP tool.

We will prove Theorem \ref{main}
by characterizing positive maps (also known as entanglement witnesses)
acting on $2\times 2$ matrices. This characterization
shall come about from the analysis of quadratic matrix
polynomials in a complex variable.

\section{\label{sec:level4}Positive maps}
We say that a linear map $\Phi : {\mathbb C}^{2 \times 2} \to {\mathbb C}^{M\times M}$
is {\it positive} when $\Phi(G) \ge 0$ whenever $G \ge 0$.
By linearity it suffices to check this condition for rank 1 matrices $G$.
Furthermore, we may assume that $G$ has the form
$$ G = \begin{pmatrix} 1 & 0 \cr 0 & 0  \end{pmatrix}\ \hbox{\rm or} \  G = \begin{pmatrix} z \cr 1 \end{pmatrix} \begin{pmatrix} \overline{z} & 1 \end{pmatrix}
, z \in {\mathbb C} .
$$
If we let $P=\Phi(E_{11})$, $Q=\Phi(E_{12})$ and $R= \Phi(E_{22})$,
where $E_{ij}$ is the $2\times 2$ matrix with a $1$ in entry $(i,j)$
and zeros elsewhere, checking $\Phi(G) \ge 0$ for all
$G$ as above amounts to checking that $P\ge 0$ and
\begin{equation}\label{quad}
|z|^2 P + zQ + \overline z Q^\dag + R \ge 0, \ \ z\in {\mathbb C} .
\end{equation}
Since $P\ge 0$ is automatically satisfied when \eqref{quad} holds (let $|z| \to \infty$) it therefore
suffices to check \eqref{quad}. In other words, we have the following lemma.

\begin{lemma} \label{pm}
The map $\Phi : {\mathbb C}^{2 \times 2} \to {\mathbb C}^{M\times M}$
is positive if and only if $P=\Phi(E_{11})$,
$Q=\Phi(E_{12})$ and $R= \Phi(E_{22})$ satisfy \eqref{quad}.
\end{lemma}

We therefore need to study quadratic matrix polynomials in a complex variable.

\section{\label{sec:level3}Quadratic matrix polynomials in complex variables}

For $M\times M$ matrices $P$, $Q$ and $R$, consider the matrix inequalities
\eqref{quad}.
As observed, \eqref{quad} implies $P \ge 0$, and clearly it also implies
$R \ge 0$. Our analysis is based on the observation
that we can eliminate arg$z$ and subsequently $|z|$ in inequality
\eqref{quad}. We do this as follows.
Write $z = r e^{j\theta}$ with $r,\theta \in {\mathbb R}$.
Then \eqref{quad} is equivalent to
\begin{equation}
\label{quad2}
r^2 P + r (e^{j\theta} Q + e^{-j\theta} Q^\dag ) + R \ge 0 ,
\ \ r,\theta \in {\mathbb R}.
\end{equation}
We now use the following
well-known fact. For completeness we provide a proof.
Useful references on Toeplitz operators include \cite[Chapter XXIII]{GGK} and
\cite{BS}.
\begin{lemma}
\label{fact}
Given is the operator-valued trigonometric polynomial $H(z):=z^{-1} S^\dag + T + zS$,
with $S$ and $T$ bounded linear operators on a Hilbert space ${\mathcal H}$.
Then $H(z) \ge 0$, $|z| = 1$, if and only if the Toeplitz operator
\begin{equation}
\label{toep3}
\begin{pmatrix} T & S^\dag & 0 & 0 & \cdots \cr
S &  T & S^\dag & 0 &\cdots \cr
0 & S &  T & S^\dag & \cdots \cr
\vdots & \ddots & \ddots & \ddots & \ddots \cr
\end{pmatrix}
\end{equation}
is positive semidefinite.
\end{lemma}

\noindent{\bf Proof.} Consider the multiplication operator $g \to
Hg$ on the Lebesgue space $L_2({\mathbb T} , {\mathcal H})$ of square
integrable Lebesgue measurable functions on the unit circle ${\mathbb T}$
with values in ${\mathcal H}$. By identifying $L_2({\mathbb T} , {\mathcal H})$
with the Hilbert space $\ell_2({\mathbb Z} , {\mathcal H})$
of square summable (in norm) sequences $(h_j)_{j=-\infty}^\infty$, $h_j \in {\mathcal H}$,
the positive semidefiniteness of $H(z)$ for all $z \in {\mathbb T}$ is equivalent
to the positive semidefiniteness of the doubly infinite Toeplitz matrix $\Lambda$
with symbol $H(z)$ (that is, $\Lambda$ is the doubly infinite
version of \eqref{toep3}). As \eqref{toep3} is the restriction of this
doubly infinite Toeplitz matrix to $\ell_2({\mathbb N} , {\mathcal H})$,
positive semidefiniteness of \eqref{toep3} follows.

Conversely, let $h= (h_j)_{j=-\infty}^\infty \in \ell_2({\mathbb Z} , {\mathcal H})$
be so that $h_j=0$, $j \ge B$. Then it follows from the positive
semidefiniteness of \eqref{toep3} that $\langle \Lambda h, h \rangle \ge 0$.
Since sequences $h$
of the above form are dense in $\ell_2({\mathbb Z} , {\mathcal H})$,
$\Lambda \ge 0$ follows. But since $H(z)$ is the symbol of this
multiplication operator $\Lambda$, it follows that $H(z) \ge 0$, $|z|=1$. $\square$

Fixing $r$ and applying Lemma \ref{fact} to \eqref{quad2},
we obtain that \eqref{quad2} holds for all $\theta$ if and
only if the following infinite block Toeplitz matrix is positive semidefinite:
\begin{equation}
\label{toep}
\begin{pmatrix} R+r^2P & rQ^\dag & 0 & 0 & \cdots \cr
rQ &  R+r^2P & rQ^\dag & 0 &\cdots \cr
0 & rQ &  R+r^2P & rQ^\dag & \cdots \cr
\vdots & \ddots & \ddots & \ddots & \ddots \cr
\end{pmatrix} \ge 0 , r \in {\mathbb R}.
\end{equation}
Notice that we may write \eqref{toep} as
\begin{equation}
\label{kln}
r^2 K + rL + N \ge 0 , \ \ r \in {\mathbb R},
\end{equation}
where $K$, $L$ and $N$ are infinite self-adjoint block Toeplitz
matrices.
Theorem 6.7 in \cite{RR} states that an operator polynomial
of even degree, $2q$ say, that is positive semidefinite for all $r \in {\mathbb R}$
allows a factorization $T(r)T(r)^\dag$, where $T(r)$ is an operator polynomial
of degree $q$. In our case we obtain that
$$ r^2 K + rL + N = (rT+S)(rT+S)^\dag .$$
This yields that
$K= TT^\dag$, $L =TS^\dag+ ST^\dag$
and $N= SS^\dag$. Consequently, if we let $X = TS^\dag - \frac{L}{2}$ we get that
$X$ is skew-adjoint ($X = -X^\dag$) and
\begin{equation}
\label{ineq}
\begin{pmatrix} K & \frac{L}{2} + X \cr \frac{L}{2}+X^\dag & N \end{pmatrix} \ge 0. \end{equation}
Thus \eqref{kln} implies the existence of $X = -X^\dag$ such that
\eqref{ineq} holds. The converse is also valid, as
$$r^2 K + rL + N =\begin{pmatrix} rI & I \end{pmatrix}
\begin{pmatrix} K & {L \over 2} + X \cr {L\over 2}-X & N \end{pmatrix}
\begin{pmatrix} rI \cr I \end{pmatrix} . $$
Applying this to our particular choice of $K, L$ and $N$ in terms of $P$,
$Q$ and $R$, we get that \eqref{quad} holds
if and only if there exists a skew-adjoint $X = (X_{ij})_{i,j\ge 0}$ so that
\begin{widetext}
\begin{equation}
\label{toep2}
\begin{pmatrix} {\begin{pmatrix} P & 0 & \cdots \cr 0 & P & \cdots \cr
\vdots & & \ddots \end{pmatrix}} &
{\begin{pmatrix} X_{00} & \frac{Q^\dag}{2} + X_{01} & X_{02} & \cdots \cr
\frac{Q}{2}+ X_{10} & X_{11} & \frac{Q^\dag}{2} + X_{12} & \cdots \cr
\vdots & & \ddots \end{pmatrix}} \cr
{\begin{pmatrix} X_{00}^\dag & \frac{Q^\dag}{2} + X_{10}^\dag & X_{20}^\dag & \cdots \cr
\frac{Q}{2}+ X_{01}^\dag & X_{11}^\dag & \frac{Q^\dag}{2} + X_{21}^\dag & \cdots \cr
\vdots & & \ddots \end{pmatrix}} & {\begin{pmatrix} R& 0 & \cdots \cr 0 & R & \cdots \cr
\vdots & & \ddots \end{pmatrix}}
\end{pmatrix} \ge 0 .
\end{equation}
Performing a permutation, we may rewrite \eqref{toep2} as
\begin{equation}
\label{Lm0}
\Lambda_0:= \begin{pmatrix} {\begin{pmatrix} P & X_{00} \cr X_{00}^\dag & R \end{pmatrix}}
& {\begin{pmatrix} 0 & \frac{Q^\dag}{2} + X_{01}\cr \frac{Q^\dag}{2} + X_{10}^\dag
& 0 \end{pmatrix}} & {\begin{pmatrix} 0 & X_{02} \cr X_{20}^\dag & 0  \end{pmatrix}}
& \cdots \cr
{\begin{pmatrix} 0 & \frac{Q}{2} + X_{10}\cr \frac{Q}{2} + X_{01}^\dag& 0\end{pmatrix}} &
 {\begin{pmatrix} P & X_{11} \cr X_{11}^\dag & R \end{pmatrix}}
& {\begin{pmatrix} 0 & \frac{Q^\dag}{2} + X_{12}\cr \frac{Q^\dag}{2} + X_{21}^\dag
& 0 \end{pmatrix}} & \cdots \cr
\vdots & \ddots & \ddots & \ddots & \end{pmatrix} \ge 0.
\end{equation}
\end{widetext}
Next we want to show that we may in fact choose
$X_{ij} = X_{i-j}$ for all $i$ and $j$. Perhaps the quickest way to do
this is by using
Banach limits (for the definition, see \cite[Section III.7]{Conway}).
Carrying out some of the ideas related to
Banach limits directly
to the current situation, we obtain the following argument.
Let $\Lambda_n$ be the
infinite principal submatrix obtained from $\Lambda_0$ by omitting the first
$n$ block rows and columns, each of size $2M$. Thus $\Lambda_n$ has
$\begin{pmatrix} P & X_{nn} \cr X_{nn}^\dag & R \end{pmatrix}$ in the top right
corner. Of course, $\Lambda_n \ge 0$ for all $n$, and the sequence
$\{ \Lambda_n \}_{n \ge 0}$ is bounded (by $\| \Lambda_0 \|$).
Consider now the bounded sequence $\{ \Xi_n \}_{n\ge 0}$ of averages
defined via
$$ \Xi_n := \frac{1}{n+1} (\Lambda_0 + \cdots + \Lambda_n ) \ge 0, \ n \in
{\mathbb N}_0. $$
This sequence has a convergent subsequence $\{ \Xi_{n_k} \}_{k\ge 0}$
in the weak operator topology. Notice that $\Xi_{n_k}$ has the same form
as $\Lambda_0$; only the operators $X_{ij}$ are different.
Therefore, its limit $\Xi_\infty$, which is positive semidefinite, must have the
same form as $\Lambda_0$ as well, with $X_{ij}$ replaced by $Y_{ij}$, say.
We claim that $Y_{ij}= Y_{i+1,j+1}$ for all $i$ and $j$. Indeed, since
$\lim_{k\to \infty} {X_{ij} \over n_k+1} = 0$ and
$\lim_{k\to \infty} {X_{i+n_k+1,j+n_k+1} \over n_k+1} = 0$,
we get that
$$ Y_{ij} - Y_{i+1,j+1} = \lim_{k\to \infty} {X_{ij} + \cdots + X_{i+n_k, j+n_k}
\over n_k+1} - $$
$$ \lim_{k\to \infty} {X_{i+1,j+1} + \cdots + X_{i+n_k+1, j+n_k+1}
\over n_k+1} = $$ $$\lim_{k\to \infty} {X_{ij} \over n_k+1} -{X_{i+n_k+1,j+n_k+1} \over n_k+1} = 0. $$
In conclusion,
we have obtained that \eqref{quad} holds if and only if there exist
$X_i = -X_{-i}^\dag$, $i \in {\mathbb N}$ so that
\begin{widetext}\begin{equation}
\label{mess}
\begin{pmatrix} {\begin{pmatrix} P & X_{0} \cr -X_{0} & R \end{pmatrix}}
& {\begin{pmatrix} 0 & \frac{Q^\dag}{2} + X_{-1}\cr \frac{Q^\dag}{2} - X_{-1}
& 0 \end{pmatrix}} & {\begin{pmatrix} 0 & X_{-2} \cr -X_{-2} & 0  \end{pmatrix}}
& \cdots \cr
{\begin{pmatrix} 0 & \frac{Q}{2} + X_{1}\cr \frac{Q}{2} - X_{1}& 0\end{pmatrix}} &
 {\begin{pmatrix} P & X_{0} \cr -X_{0} & R \end{pmatrix}}
& {\begin{pmatrix} 0 & \frac{Q^\dag}{2} + X_{-1}\cr \frac{Q^\dag}{2} - X_{-1}
& 0 \end{pmatrix}} & \cdots \cr
\vdots & \ddots & \ddots & \ddots & \end{pmatrix} \ge 0.
\end{equation}
\end{widetext}

Let us denote the principal $(n+1) \times (n+1)$ block submatrix
of \eqref{mess} by ${\mathcal D}(X_0, \ldots , X_n ; P, Q, R )$.
Then we obtain the following result.

\begin{proposition} \label{qq}
The matrices $P,Q$ and $R$ satisfy \eqref{quad} if and only if
there exist $ X_i = -X_{-i}^\dag$, $i =0,1,\ldots $ so that
${\mathcal D}(X_0, \ldots , X_n ; P, Q, R ) \ge 0$ for all $n$.
\end{proposition}

Let ${\mathcal C}_n$ denote the cone of all matrices $\sigma =
\begin{pmatrix} P &
Q \cr Q^\dag & R \end{pmatrix}$ so that there exist $ X_i = -X_{-i}^\dag$, $i =0, \ldots , n$, so that ${\mathcal D}(X_0, \ldots , X_n ; P, Q, R ) \ge 0$.
Using this notation we obtain the following.

 \begin{proposition} \label{qq2}
The matrices $P,Q$ and $R$ satisfy \eqref{quad} if and only if
$$\sigma =
\begin{pmatrix} P &
Q \cr Q^\dag & R \end{pmatrix} \in \cap_{n\ge 1} {\mathcal C}_n . $$
\end{proposition}

Combining Lemma \ref{pm} and Proposition \ref{qq2} we now obtain the following description
of positive maps acting ${\mathbb C}^{2 \times 2} \to
{\mathbb C}^{M\times M}$.

\begin{proposition} \label{qq3}
The map $\Phi : {\mathbb C}^{2 \times 2} \to {\mathbb C}^{M\times M}$
is positive if and only if
$$\begin{pmatrix} \Phi(E_{11}) &
\Phi(E_{12}) \cr \Phi(E_{21} & \Phi(E_{22})
\end{pmatrix} \in \cap_{n\ge 1} {\mathcal C}_n . $$
\end{proposition}

Next we need duality to get back to the separability problem. We first
prove the following auxiliary result.

\begin{lemma} \label{duality} The cones ${\mathcal A}_n$ and
${\mathcal C}_n$ are one another's dual. \end{lemma}

\noindent {\bf Proof.} Let $\rho \in {\mathcal A}_n$. In
order to show that $\rho \in {\mathcal C}_n^*$ we need to show that
trace$(\rho \sigma )\ge 0$ for all $\sigma \in {\mathcal C}_n$. Thus, let
$ \sigma \in {\mathcal C}_n$, and let $X_i = -X_{-i}^\dag$, $i=0,
\ldots , n$, be so that $D:= {\mathcal D}(X_0, \ldots , X_n ; P, Q, R ) \ge 0$.
Since $\rho \in {\mathcal A}_n$ there exists a $\Gamma
\in {\mathcal G} (\rho ; n )$. It is now
straightforward to check that trace$(\rho \sigma ) = \hbox{\rm trace} (\Gamma
D)$, and since $D$ and $\Gamma$ are positive semidefinite, trace$(\rho \sigma )
\ge 0$ follows. This shows that ${\mathcal A}_n \subset {\mathcal C}_n^*$.

For the converse inclusion, observe that when
$\rho = \begin{pmatrix} A & B \cr B^\dag & D \end{pmatrix}
\not \in {\mathcal A}_n $, then for
all $\Gamma_{ij}$ satisfying (i)-(v) we have that
$(\Gamma_{ij})_{i,j=0}^n \not\ge 0$. Notice that
the block matrices $(\Gamma_{ij})_{i,j=0}^n$ satisfying (i)-(v)
describe a affine space, which we may denote as
$G+ {\mathcal L}$ where $G$ is a fixed matrix and ${\mathcal L}$
is a linear subspace (described by all matrices satisfying (i)-(v) with
$A=D=B=0$).
This affine space $G+ {\mathcal L}$ is separated from the cone PSD of positive semidefinites.
Thus by the Hahn-Banach theorem and the selfduality of
PSD there exist a positive semidefinite
$W$ so that $\hbox{\rm trace}(W (G+L)) <0$
for all $L \in {\mathcal L}$. But then $W\in {\mathcal L}^\perp$
and thus $W$
is of the form
${\mathcal D}(X_0, \ldots , X_n ; P, Q, R )$. Let $\sigma
= \begin{pmatrix} P & Q \cr Q^\dag & R \end{pmatrix}$, which
belongs to ${\mathcal C}_n$. As trace$( \sigma \rho ) =
\hbox{\rm trace} (W G) < 0$, it follows that
$ \rho \not \in {\mathcal C}_n^*$. Thus ${\mathcal A}_n = {\mathcal C}_n^*$.

Since ${\mathcal C}_n$ is closed, ${\mathcal A}_n^* = {\mathcal C}_n^{**}
={\mathcal C}_n$ follows also. $\square$

We are now ready to prove the main result.

\noindent {\bf Proof of Theorem \ref{main}} By \cite{HHH} we have that
the cone of $2 \times M$ separable matrices has the dual
$$ \{ \begin{pmatrix} \Phi(E_{11}) &
\Phi(E_{12}) \cr \Phi(E_{21} & \Phi(E_{22})
\end{pmatrix} : \Phi : {\mathbb C}^{2 \times 2} \to {\mathbb C}^{M\times M}
\ \hbox{\rm is positive } \} , $$
which by Proposition \ref{qq3} equals $\cap_{n=1}^\infty {\mathcal C}_n$.
The cone of $2\times M$ separable matrices therefore equals
$$(\cap_{n=1}^\infty {\mathcal C}_n)^* = \overline {\cup_{n=1}^\infty
{\mathcal C}_n^*} = \overline{{\cup_{n=1}^\infty {\mathcal A}_n}}
. $$ In the last step we used Lemma \ref{duality}. $\square$

\begin{acknowledgments}
The research was done while the author was
visiting the Electrical Engineering department at the Katholieke
Universiteit Leuven
and the
Department of Mathematical Engineering at Universit\'e Catholique
de Louvain.
The author gratefully acknowledges the hospitality of
his hosts Bart de Moor (KUL)
and Paul Van Dooren (UCL).
The author also
thanks Paul Van Dooren and Yurii Nesterov for useful
discussions on the topic of quadratic matrix functions,
Jeroen Dehaene for his comments on the manuscript, and the
referees for their helpful comments.
Finally, the research was
supported by the College of William and Mary (Faculty Research
Assignment grant) and
NSF grant DMS 9988579.

\end{acknowledgments}

\def\polhk#1{\setbox0=\hbox{#1}{\ooalign{\hidewidth
  \lower1.5ex\hbox{`}\hidewidth\crcr\unhbox0}}}

\end{document}